\documentclass[epj]{svjour}
\usepackage{graphics}
\usepackage{bm}
\usepackage{amsfonts}
\usepackage{latexsym}
\usepackage[latin1]{inputenc}
\usepackage{graphicx}
\usepackage{amsmath}
\usepackage{palatino}
\usepackage{mathpazo}
\usepackage{textcomp}
\usepackage{float}
\usepackage{booktabs}
\usepackage{dcolumn}
\usepackage{ragged2e}
\usepackage{hyperref}
\hypersetup{colorlinks,citecolor=blue}
\usepackage{amsmath}
\usepackage{xcolor}
\usepackage{orcidlink}
\usepackage[caption=false]{subfig}
\usepackage{commath}
\begin{document}
\title{Cosmological bouncing scenarios in symmetric teleparallel gravity}
%\subtitle{Do you have a subtitle?\\ If so, write it here}
\author{Sanjay Mandal\inst{1}\orcidlink{0000-0003-2570-2335}, N. Myrzakulov\inst{2}\inst{,3}\orcidlink{0000-0001-8691-9939}, P.K. Sahoo\inst{1}\orcidlink{0000-0003-2130-8832} \and R. Myrzakulov \inst{2}\inst{,3}\orcidlink{0000-0002-5274-0815}}% etc
% \thanks is optional - remove next line if not needed
%\thanks{\emph{E-mail addresses:} Sanjay Mandal:sanjaymandal960@gmail.com, N. Myrzakulov:nmyrzakulov@gmail.com, P.K. Sahoo:pksahoo@hyderabad.bits-pilani.ac.in, R. Myrzakulov:rmyrzakulov@gmail.com}%
                    % Do not remove
%
%%\offprints{}          % Insert a name or remove this line
%
\institute{Department of Mathematics, Birla Institute of Technology and
Science-Pilani, Hyderabad Campus, Hyderabad-500078, India. 
\and Ratbay Myrzakulov Eurasian International Centre for Theoretical Physics, Nur-Sultan 010009, Kazakhstan \and Gumilyov Eurasian  National University, Nur-Sultan 010008, Kazakhstan}

\date{Received: 1st June 2021 / Revised version: 26th June 2021}
% The correct dates will be entered by Springer
%
\abstract{
Symmetric Teleparallel Gravity is an exceptional theory of gravity that is consistent with the vanishing affine connection. This theory is an alternative and a simpler geometrical formulation of general relativity, where the non-metricity $Q$ drives the gravitational interaction. Our interest lies in exploring the cosmological bouncing scenarios in a flat Friedmann-Lima\^itre-Robertson-Walker (FLRW) spacetime within this framework. We explore bouncing scenarios with two different Lagrangian forms of $f(Q)$ such as a linearly and non-linearly dependence on $Q$. We have successfully examined all the energy conditions and stability analysis for both models to present a matter bounce.
 % end of PACS codes
} %
\titlerunning{Cosmological bouncing scenarios in symmetric teleparallel gravity} 
\authorrunning{S. Mandal et al.}
\maketitle
\section{Introduction}

According to recent astrophysical observations like Supernovae Ia \cite{SN1}, large scale structure \cite{LSS} with the baryon acoustic oscillations \cite{Eisenstein:2005su}, cosmic microwave background radiation \cite{WMAP,Komatsu:2008hk,Komatsu:2010fb}, and weak lensing \cite{Jain:2003tba}, the current evolution expansion of the universe is accelerating.  As suggested by observations Universe is flat, homogeneous, isotropic, and corresponds to Friedman-Robertson-Walker space-time. The current status of the universe motivated the research community to go beyond Einstein's general relativity (GR) to explore the candidates responsible for it. In GR, the addition of cosmological constant in field equations helps us to understand the unknown form of energy, but it faces some issues such as the fine-tuning problem, coincidence problem; its effects are only observed at cosmological scales instead of Planck scales \cite{carlip/2019}. Therefore, several alternative modified theories of gravity have been introduced to address the issues with GR in the literature.

At the present time, we have two known procedures to explain the late-time cosmic acceleration. One is the introduction of the so-called dark energy (DE) with negative pressure in Einstein's general theory of relativity \cite{DE}. After the discovery of late-time cosmic acceleration numerous DE models have been proposed, such as phantom \cite{Phantom}, quintessence\cite{Quint} , $k$-essence\cite{kessence} , $f$-essence\cite{fessence} as well as tachyon \cite{Tach}  and so on. The other is the modification of theories of gravity on large distances. During the last years, it has been tested that modified theories of gravity are quite successful in the description of the early-time like inflation and late-time acceleration like dark energy of our Universe. 

In the last few years, modified theories of gravity have been proposed as extensions of Einstein's general relativity theory. Their main motivation is to explain the latest astrophysical data on dark energy and dark matter. The investigations of general relativity at small scales have already given important results, but its study at large scales is challenging because recent observational data will provide important information on the validity of these modified theories of gravity. In particular, $f(R)$ gravity, $f(T)$ gravity extensively studied as geometrical description of gravity. 

In $f(R)$ theories of gravity, the scalar curvature $R$ in the Einstein-Hilbert action is changed to an arbitrary function $f(R)$ and described by a Levi-Civita connection \cite{fr}. This connection describes a spacetime with zero torsion and nonmetricity but with nonvanishing curvature. The $f(T)$ gravity is a generalization of the teleparallel equivalent of general relativity (TEGR), where $T$ is the torsion scalar made up of the Weitzenbock connection in which a spacetime describes with zero curvature and nonmetricity but with nonvanishing torsion \cite{ft}. 

Recently, authors \cite{STG,CGR} proposed symmetric teleparallel gravity (STG) or $f(Q)$ theory of gravity based on nonmetricity scalar $Q$. This is a new geometric interpretation of space-time in affine connection, where curvature and torsion vanish. There are several mathematical and physical reasons which motivate the consideration of nonmetricity in the context of gravitational theories. For instance, nonmetricity scalar is assuming as a measure for the violation of local Lorentz invariance, which has been attracting some interest.
In \cite{STGDS} explored cosmological applications in STG and investigated dynamical system method. Possibility of constructing conformal theories of gravity in the STG examined in \cite{STGCG}. Authors \cite{STGGW,GWSTG} studied the propagation velocity of the gravitational waves around Minkowski spacetime and their potential polarizations in a general class of STG. Moreover, in \cite{GWSTG},  authors have investigated the possibility of GWs exactly for STG and their modifications with two extensions of STG, namely the perturbed versions of the generalized irreducible decomposition of the STG which emerges from nonmetricity scalar $Q$, as well as the other natural generalization of the theory, namely $f(Q)$ gravity. They found that polarization modes turn out to be identical to those of GR. This means that GW polarization tests cannot distinguish between GR and $f(Q)$ theories of gravity, while the linear irreducible form of STG theory does produce distinct results that would emerge in GW observations. Also, the authors investigated GWs in theories with a nonminimal coupling between $f(Q)$ and a scalar field.  Additionally, authors examined GW in the extension of STG such as $f(Q, B)$ gravity, where the $Q$ nonmetricity scalar and $B$ boundary term. Cosmographic parameters in $f(Q)$ gravity analized in \cite {CPQ}. In \cite {WGQ} studied the traversable wormhole geometries in $f(Q)$ gravity and constructed geometries for three different cases. 

In order to give characteristics that describe some realistic matter distribution, certain conditions must be imposed on the energy-momentum tensor known as energy conditions originate from Raychaudhuri equations with the requirement that energy density is positive. The energy conditions as null (NEC), weak (WEC), dominant (DEC), and strong (SEC) energy conditions are the four fundamental conditions that are often required in the proofs of various important theorems about black holes, such as the laws of black hole thermodynamics in the general theory of relativity. These conditions have investigated in various modified theories of gravity such as \cite{EC1,EC2,EC3,EC4,EC7,EC5,EC6,Ovgun/2019}.

One alternative theory for inflation is the bouncing cosmology in general. This theory constructed as following: scale factor reaches a minimum but finite value in inflationary scenario. Also the bouncing universe assuming as a new idea, suggested to resolves the singularity problem in big bang cosmology. In this interpretation, the initial phase of the universe transfer to a late-time accelerated expansion phase in which point the Hubble parameter, $H(t)$, transit from $H(t)<0$ to $H(t)>0$ and also in bounce point, we have $H(t)=0$. In the literature, an increasing number of works concerning potential viable bouncing cosmologies have been examined. Different types of bounce like ekpyrosis bounce and superbounce have also been attracting interest in modified theories framework. For instance in the literature with seminal works in which bounce scenarios were explored for $f(R)$, $f(T)$ and $f(G)$ gravity theories \cite{BounceR,BounceT,BounceG} and (for a recent review, see \cite{BounceRev}). Recently, one of the interesting work was done by F. Bajardi er atl., \cite{Bajardi/2020} in the framework of $f(Q)$ gravity, there they used the reduction method to constraint the functional form of $f(Q)$ and explore the bouncing cosmological models. In this work, we are aiming to explore the cosmological bouncing scenarios in symmetric teleparallel gravity (i.e. $f(Q)$ gravity).

This manuscript is organized as follows: in Sec \ref{II}, we briefly discuss the gravitational formalism for $f(Q)$ gravity. In Section \ref{III}, we present the energy conditions. Then, we discuss different types of bouncing scenarios and also about a particular bouncing solution, which we have used for further study in Sec. \ref{IV}. In Sec. \ref{V}, we explore two bouncing models with their energy conditions. We adopt the perturbation analysis to examine the stability of bouncing scenarios in Sec. \ref{VI}. Finally, we conclude with our outcomes in Sec. \ref{VII}.

\section{Motion Equations in $f(Q)$ gravity}\label{II}

Let us consider the action for $f(Q)$ gravity given by \cite{CGR}
\begin{equation}
\label{1}
\mathcal{S}=\int \frac{1}{2}\,f(Q)\,\sqrt{-g}\,d^4x+\int \mathcal{L}_m\,\sqrt{-g}\,d^4x\,,
\end{equation}
where $f(Q)$ is a general function of the $Q$, $\mathcal{L}_m$ is the matter Lagrangian density and $g$ is the determinant of the metric $g_{\mu\nu}$.\\
The nonmetricity tensor and its traces are such that
\begin{equation}
\label{2}
Q_{\gamma\mu\nu}=\nabla_{\gamma}g_{\mu\nu}\,,
\end{equation}
\begin{equation}
\label{3}
Q_{\gamma}={{Q_{\gamma}}^{\mu}}_{\mu}\,, \qquad \widetilde{Q}_{\gamma}={Q^{\mu}}_{\gamma\mu}\,.
\end{equation}
Moreover, the superpotential as a function of nonmetricity tensor is given by
\begin{equation}
\label{4}
4{P^{\gamma}}_{\mu\nu}=-{Q^{\gamma}}_{\mu\nu}+2Q_{({\mu^{^{\gamma}}}{\nu})}-Q^{\gamma}g_{\mu\nu}-\widetilde{Q}^{\gamma}g_{\mu\nu}-\delta^{\gamma}_{{(\gamma^{^{Q}}}\nu)}\,,
\end{equation}
where the trace of nonmetricity tensor \cite{CGR} has the form
\begin{equation}
\label{5}
Q=-Q_{\gamma\mu\nu}P^{\gamma\mu\nu}\,.
\end{equation}
Another relevant ingredient for our approach is the energy-momentum tensor for the matter, whose definition is
\begin{equation}
\label{6}
T_{\mu\nu}=-\frac{2}{\sqrt{-g}}\frac{\delta(\sqrt{-g}\mathcal{L}_m)}{\delta g^{\mu\nu}}\,.
\end{equation}
Taking the variation of action \eqref{1} with respect to metric tensor, one can find the field equations
%\begin{widetext}
\begin{equation}
\label{7}
\frac{2}{\sqrt{-g}}\nabla_{\gamma}\left( \sqrt{-g}f_Q {P^{\gamma}}_{\mu\nu}\right)+\frac{1}{2}g_{\mu\nu}f
+f_Q\left(P_{\mu\gamma i}{Q_{\nu}}^{\gamma i}-2Q_{\gamma i \mu}{P^{\gamma i}}_{\nu} \right)=-T_{\mu\nu}\,,
\end{equation}
%\end{widetext}
where $f_Q=\frac{df}{dQ}$. Besides, we can also take the variation of \eqref{1} with respect to the connection, yielding to 
\begin{equation}\label{8}
\nabla_{\mu}\nabla_{\gamma}\left( \sqrt{-g}f_Q {P^{\gamma}}_{\mu\nu}\right)=0\,.
\end{equation}
Here we are going to consider the standard Friedmann-Lema\^{\i}tre-Robertson-Walker (FLRW) line element, which is explicit written as
\begin{equation}
\label{9}
ds^2=-dt^2+a^2(t)\delta_{\mu\nu}dx^{\mu}dx^{\nu}\,,
\end{equation}
where $a(t)$ is the scale factor of the Universe. The previous line element enable us to write the trace of the nonmetricity tensor as
\begin{align*}
Q=6H^2\,.
\end{align*} 
Now, let us take the energy-momentum tensor for a perfect fluid, or
\begin{equation}
\label{10}
T_{\mu\nu}=(p+\rho)u_{\mu}u_{\nu}+pg_{\mu\nu}\,,
\end{equation}
where $p$ represents the pressure and $\rho$ represents the energy density. Therefore, by substituting \eqref{9}, and \eqref{10} in \eqref{7} one can find 
\begin{equation}
\label{11}
3H^2=\frac{1}{2f_Q}\left(-\rho+\frac{f}{2}\right)\,,
\end{equation}
\begin{equation}
\label{12}
\dot{H}+3H^2+\frac{\dot{f_Q}}{f_Q}H=\frac{1}{2f_Q}\left(p+\frac{f}{2}\right)\,,
\end{equation}
as the modified Friedmann equations for $f(Q)$ gravity. Here dot $(^.)$ represents one derivative with respect to time.
% As its known, the energy-momentum conservation implies in the following equation for the perfect fluid
%\begin{equation}
%\label{13}
%\dot{\rho}+3H(\rho+p)=0\,.
%\end{equation}
The modified Friedmann equations enable us to write the density and the pressure for the Universe as
\begin{equation}
\label{14}
\rho=\frac{f}{2}-6H^2f_Q\,,
\end{equation}
\begin{equation}
\label{15}
p=\left(\dot{H}+3H^2+\frac{\dot{f_Q}}{f_Q}H\right)2f_Q-\frac{f}{2}\,.
\end{equation}

In analogy with GR, we can rewrite Eq.\eqref{11}, \eqref{12} as
\begin{equation}
3H^2=-\frac{1}{2}\tilde{\rho}\,,
\end{equation}
\begin{equation}
\dot{H}+3H^2=\frac{\tilde{p}}{2}\,.
\end{equation}
where
\begin{equation}
\label{13a}
\tilde{\rho}=\frac{1}{f_Q}\left(\rho-\frac{f}{2}\right)\,,
\end{equation}
\begin{equation}
\label{14a}
\tilde{p}=-2\,\frac{\dot{f_Q}}{f_Q}\,H+\frac{1}{f_Q}\,\left(p+\frac{f}{2}\right)\,.
\end{equation}
The previous equations are going to be components of a modified energy-momentum tensor $\tilde{T}_{\,\mu\nu}$, embedding the dependence on the trace of the nonmetricity tensor.
\section{Energy conditions}\label{III}

The energy conditions (ECs) are the essential tools to understand the geodesics of the Universe. Such conditions can be derived from the well-known Raychaudhury equations, whose forms are \cite{Poisson/2004}
\begin{equation}
\label{16}
\frac{d\theta}{d\tau}=-\frac{1}{3}\theta^2-\sigma_{\mu\nu}\sigma^{\mu\nu}+\omega_{\mu\nu}\omega^{\mu\nu}-R_{\mu\nu}u^{\mu}u^{\nu}\,,
\end{equation}
\begin{equation}
\label{17}
\frac{d\theta}{d\tau}=-\frac{1}{2}\theta^2-\sigma_{\mu\nu}\sigma^{\mu\nu}+\omega_{\mu\nu}\omega^{\mu\nu}-R_{\mu\nu}n^{\mu}n^{\nu}\,,
\end{equation}
where $\theta$ is the expansion factor, $n^{\mu}$ is the null vector, and $\sigma^{\mu\nu}$ and $\omega_{\mu\nu}$ are, respectively, the shear and the rotation associated with the vector field $u^{\mu}$. For attractive gravity, equations \eqref{16}, and \eqref{17} satisfy the following conditions
\begin{align}
\label{18}
R_{\mu\nu}u^{\mu}u^{\nu}\geq0\,,\\
 R_{\mu\nu}n^{\mu}n^{\nu}\geq0\,.
\end{align}
 Therefore, if we are working with a perfect fluid matter distribution, the energy conditions recovered from standard GR are
\begin{itemize}
\item Strong energy conditions (SEC) if  $\tilde{\rho}+3\tilde{p}\geq 0\,$;

\item Weak energy conditions (WEC) if  $\tilde{\rho}\geq 0, \tilde{\rho}+\tilde{p}\geq 0\,$;

\item Null energy condition (NEC) if  $\tilde{\rho}+\tilde{p}\geq 0\,$;

\item Dominant energy conditions (DEC) if $\tilde{\rho}\geq 0, |\tilde{p}|\leq \rho\,$.
\end{itemize}

Taking Eqs.  $(\ref{13a})$ and $(\ref{14a})$  into WEC, NEC and DEC constraints, we are able to prove that

\begin{itemize}
\item Weak energy conditions (WEC) if  $\rho\geq 0, \rho+p\geq 0\,$;

\item Null energy condition (NEC) if  $\rho+p\geq 0\,$;

\item Dominant energy conditions (DEC) if $\rho\geq 0, |p|\leq \rho\,$.
\end{itemize}
corroborating with the work from Capozziello et al.\cite{Capozziello/2018} and Mandal et al. \cite{EC5}. In the case of SEC condition, we yield to the constraint
\begin{equation}\label{15a}
\rho+3\,p-6\,\dot{f_Q}\,H+f \geq 0\,.
\end{equation}

\section{Bouncing Solution}\label{IV}
The singularity problem in the standard model of cosmology and Inflection theory is debatable for a long time since it was not clear if this particular state was an inherent trace of the universe or just a consequence of the high degree of symmetry the model \cite{Novello/2008}. However, for this issue, finite singularity might give us some answers.  Regarding this, an extensive study was done by the researchers based on some of the bouncing scenarios such as

 (I)\textit{Symmetric bounce:} This type of solution was first introduced by \cite{Cai/2012} to generate non-singular bouncing. This bouncing model needs to satisfy the other cosmological behaviors; otherwise, it suffers the problems related to the Hubble horizon. And, the cosmological parameters obey the following behaviors; for $t$ tends to $\infty$, scale factor, Hubble parameter, energy density, and pressure diverges.

(II) \textit{Super-bounce:} This type of model is used to construct a universe that collapses at a point $t_b$ (bouncing point) and re-birth through a Big Bang without a singularity \cite{Koehn/2014}. The cosmological parameters may follow the following; for $t$ tends to $t_b$, the scale factor is constant, Hubble parameter, energy density, and pressure diverge.

(III) \textit{Oscillatory cosmology:} This type of model describes the cyclic universe, i.e., it continues its expansion and contraction after a specific interval of time \cite{Novello/2008}. Here are certain rules; for $t$ tends to $kt_b (k\in \mathbb{Z}-{0})$, scale factor tends to constant, Hubble parameter, energy density, and pressure diverse.

(IV) \textit{Matter bounce:} This type of model was studied during the early stage evolution of the universe and derived from loop quantum cosmology \cite{Singh/2006}. For $t$ tends to $t_b$, scale factor, Hubble parameter, energy density, and pressure is finite.

(V) \textit{Past/future singularities and Little Rip cosmology:}  This type of model deals with the universe's past and future singularities problem. It is generally specified into four categories: big-rip, little-rip, sudden, and big-freeze singularities \cite{Nojiri/2005}.

In this work, we are looking to discuss some bouncing cosmological models in a nobel modified theory of gravity called $f(Q)$ gravity. To proceed further. we have considered the cosmic scale factor $a(t)$ is given by \cite{Klinkhamer/2019}
\begin{equation}
\label{16}
a(t)=\left(\frac{b^2+t^2}{b^2+t_0^2}\right)^{1/4},
\end{equation}
where $b$ is an arbitrary constant and $t_0>0$.

\begin{figure}[H]
\centering
\includegraphics[scale=0.4]{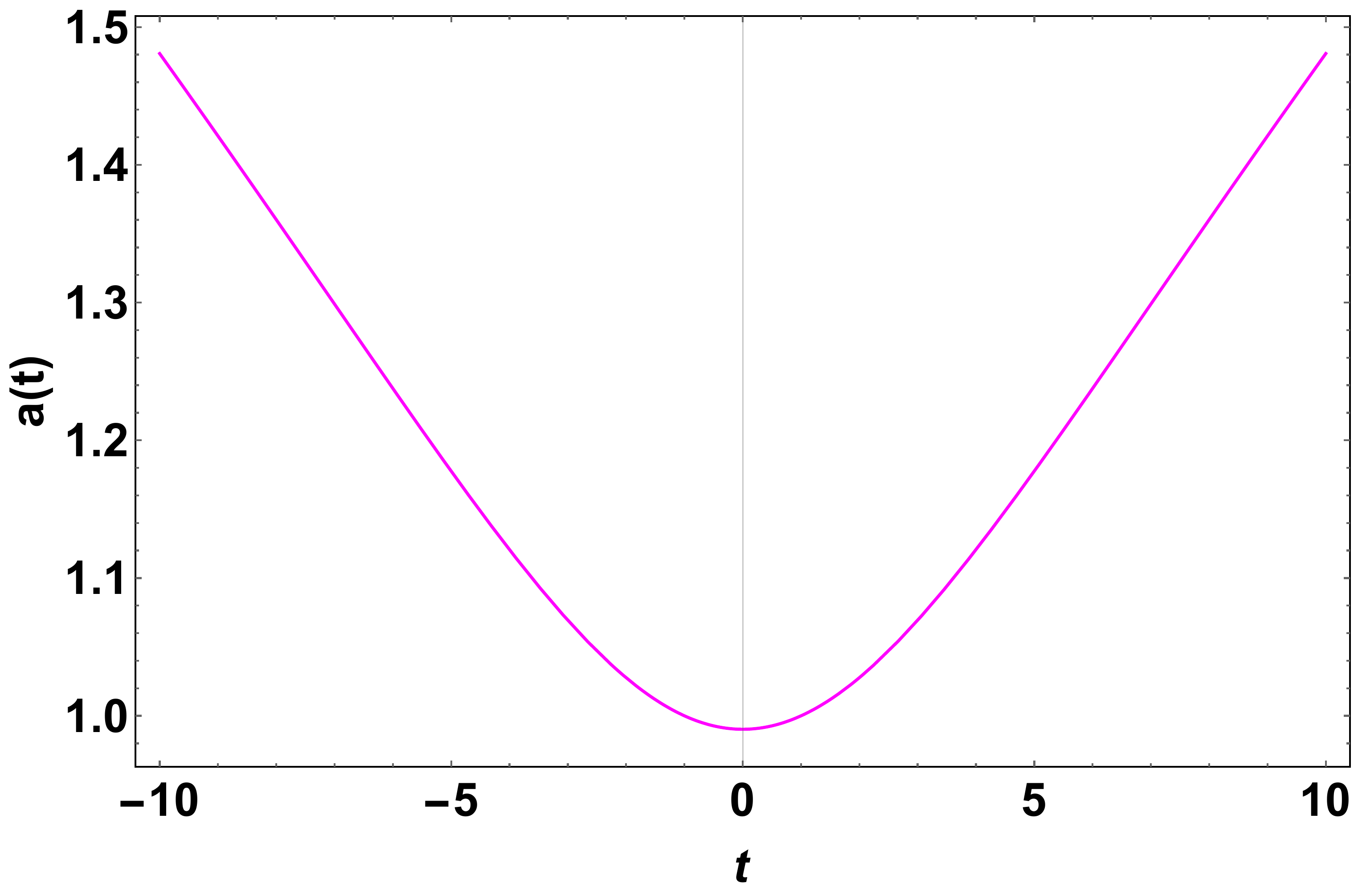}
\caption{Profile of scale factor $a(t)$ for $b=5$ and $t_0=1$.}
\label{f1}
\end{figure}

\begin{figure}[H]
\centering
\includegraphics[scale=0.4]{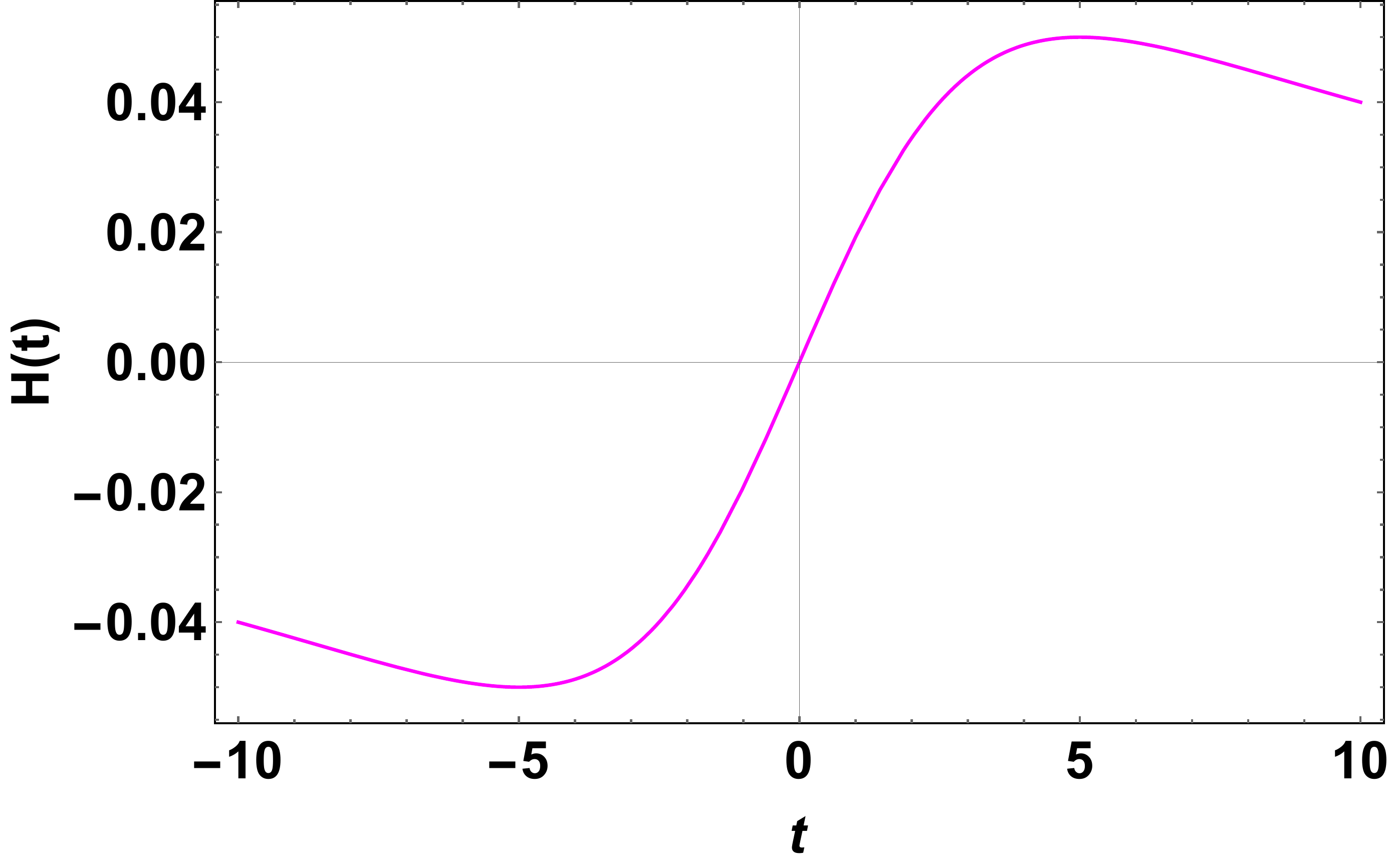}
\caption{Profile of Hubble parameter $H(t)$ for $b=5$ and $t_0=1$.}
\label{f2}
\end{figure}

The profile of the bouncing solution and its derivative are shown in Fig. \ref{f1} and \ref{f2}, respectively. From those profiles, one can see that the Hubble parameter lies in the negative range for $t<0$ and in the positive range for $t>0$. This result suggests that the universe transit from a contraction phase to an expansion phase. Moreover, for a non-zero scale factor at $t=0$, its derivative reduces to zero. One can also see that the scale factor evolved symmetrically around the bouncing point in Fig. \ref{f1}. Here, the standard units of $H$ as km/s/Mpc and $t$ as Gyr are considered.
\section{Cosmological Models}\label{V}

In this section, we discuss the presumed bouncing solution through realistic cosmological models. To do that, we discussed two cosmological models by considering two different Lagrangian $f(Q)$. For the first model, we discuss the model with linear Lagrangian, and for the second model, we assume a non-linear Lagrangian $f(Q)$. These models are discussed in the following sub-sections. 
\subsection{Model 1: $f(Q)=\alpha Q$}
 In this subsection, we consider a linear Lagrangian $f(Q)$ i.e., $f(Q)=\alpha Q$, here $\alpha$ is an arbitrary constant. Studying this linear model helps us to compare the results with the fundamental theory of gravitational interaction. For this case, the energy-momentum components are reads as follows;

\begin{equation}
\label{17}
\rho=-\frac{3 \alpha  t^2}{4 \left(b^2+t^2\right)^2}
\end{equation}
\begin{equation}
\label{18}
\textbf{WEC : }\,\ \rho+p= \frac{\alpha  (b-t) (b+t)}{\left(b^2+t^2\right)^2}
\end{equation}
\begin{equation}
\label{19}
\textbf{DEC : }\,\ \rho-p= -\frac{\alpha  \left(2 b^2+t^2\right)}{2 \left(b^2+t^2\right)^2}
\end{equation}
\begin{equation}
\label{20}
\textbf{SEC : }\,\ \rho+3\,p-6\,\dot{f_Q}\,H+f= \frac{3 \alpha  b^2}{\left(b^2+t^2\right)^2}
\end{equation}

\begin{figure}[H]
\centering
\includegraphics[scale=0.4]{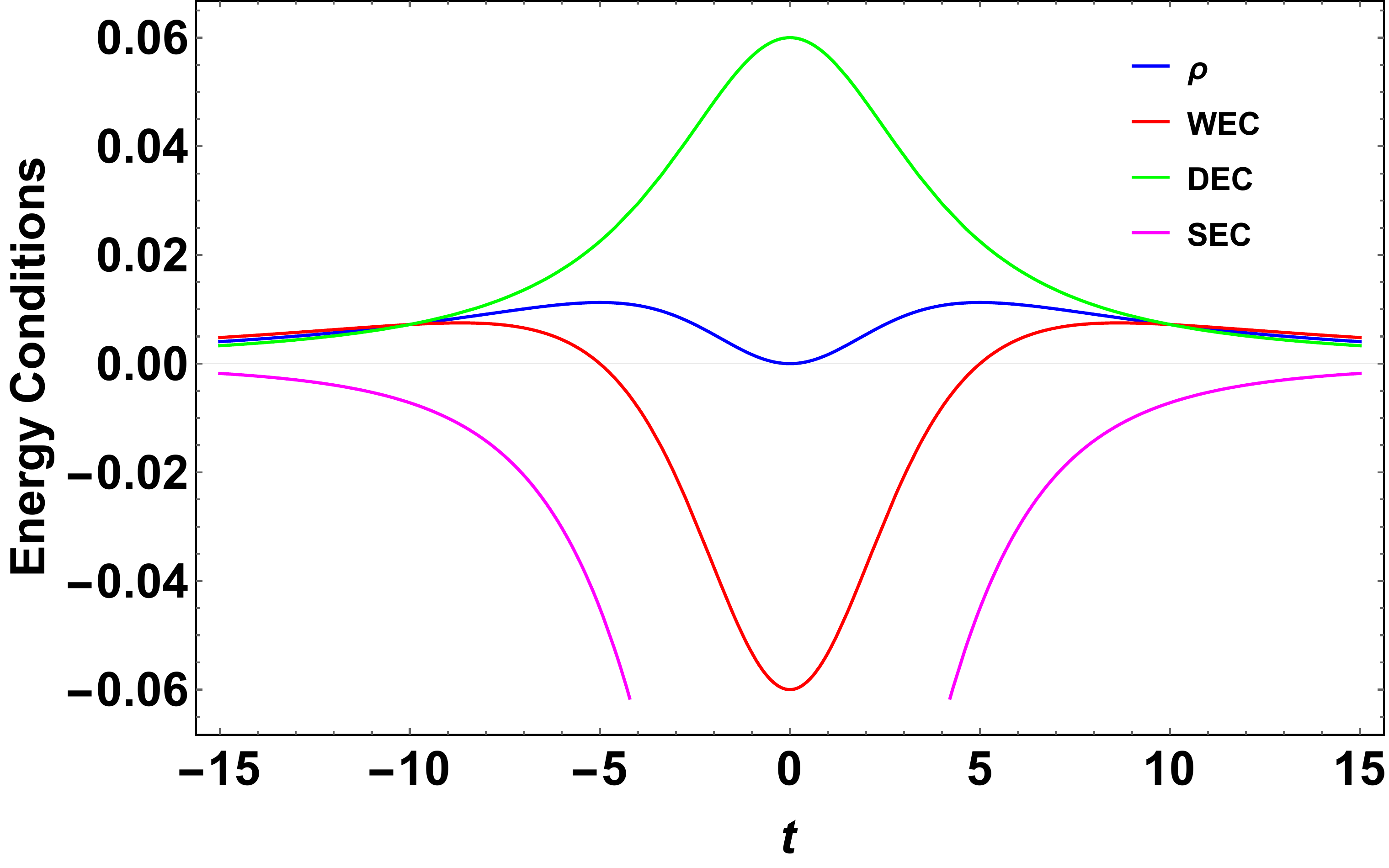}
\caption{Profiles of energy conditions w.r.t. cosmic time $t$ for $\alpha=-1.5$ and $b=5$ (for model-1).}
\label{f3}
\end{figure}

All of the above expression for energy conditions are depend on the model parameters. Nevertheless, one can not chose the values of parameters randomly, which may violate the energy conditions. Moreover, we know that energy density $\rho$ should be positive throughout the evolution of the universe. Therefore, using this condition $\rho \geq 0$, we can constrain all the parameters used to discuss our models. Therefore, for first model, we have found that for $\alpha< 0$, the energy density will be positive everywhere. The profiles of all energy conditions are depicted in Fig. \ref{f3}, for a particular value of $\alpha$ from its' bounded range. From that figure, one can see that the non-singular bounce happens at $t=0$. The energy density contracted from past evolution to $t=0$ and again expanded from that point. Besides, NEC and SEC violated near the bouncing point. Also, we have verified that the NEC and SEC are non-singular at the bouncing point, which removes the singularity issue with the early description and represents a non-singular bouncing universe.  Moreover, this model shows the late-time acceleration of the universe.

The equation of State parameter $\omega$ for this model can be written as 

\begin{equation}
\label{20a}
\omega=\frac{p}{\rho}=-1-\frac{4 \left(b^2-t^2\right)}{3 t^2}
\end{equation}

\begin{figure}[H]
\centering
\includegraphics[scale=0.4]{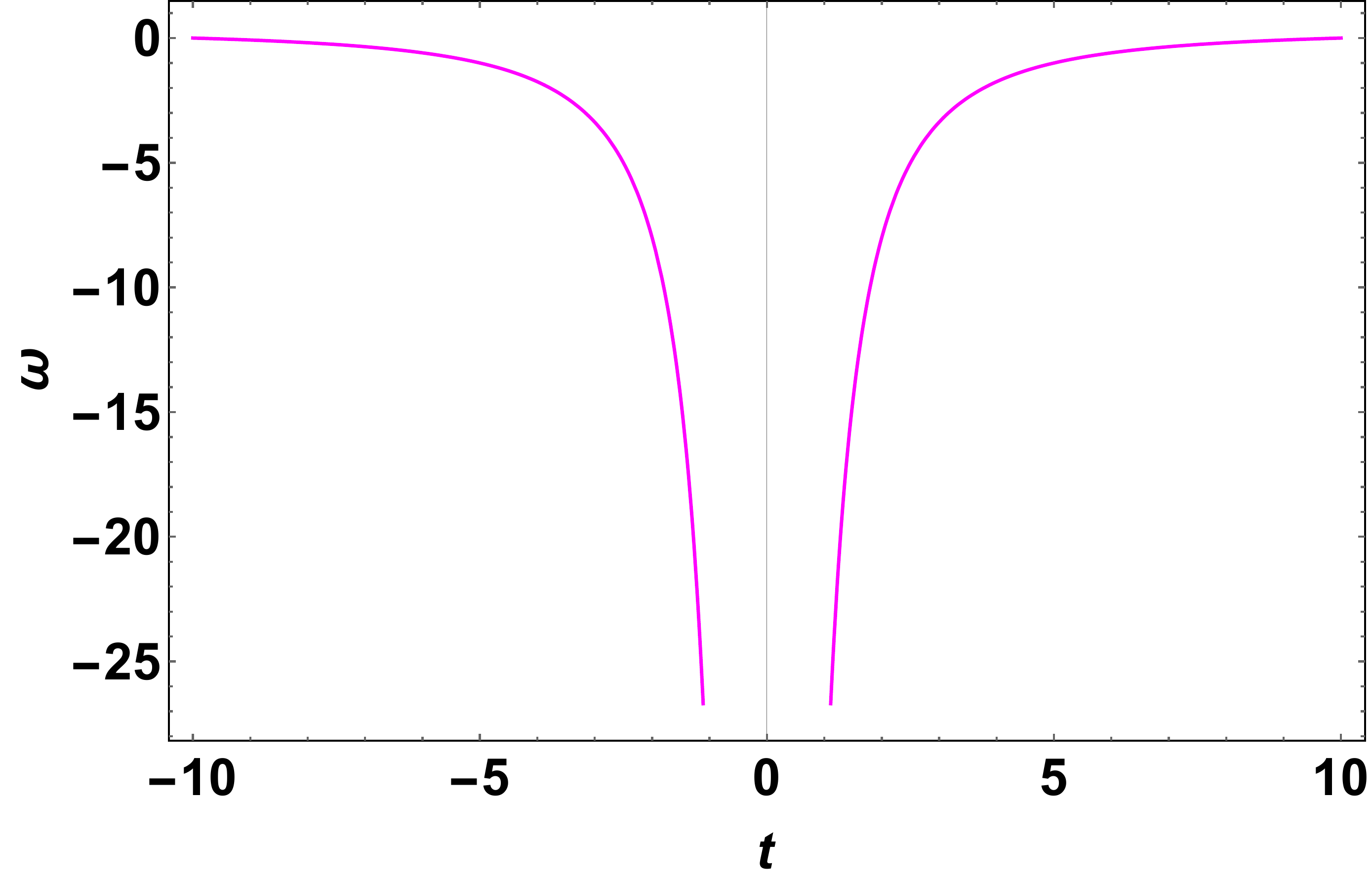}
\caption{Profile of equation of State parameter $\omega$ w.r.t. cosmic time $t$ for $\alpha=-1.5$ and $b=5$ (for model-1).}
\label{f4}
\end{figure}

\begin{figure}[H]
\centering
\includegraphics[scale=0.4]{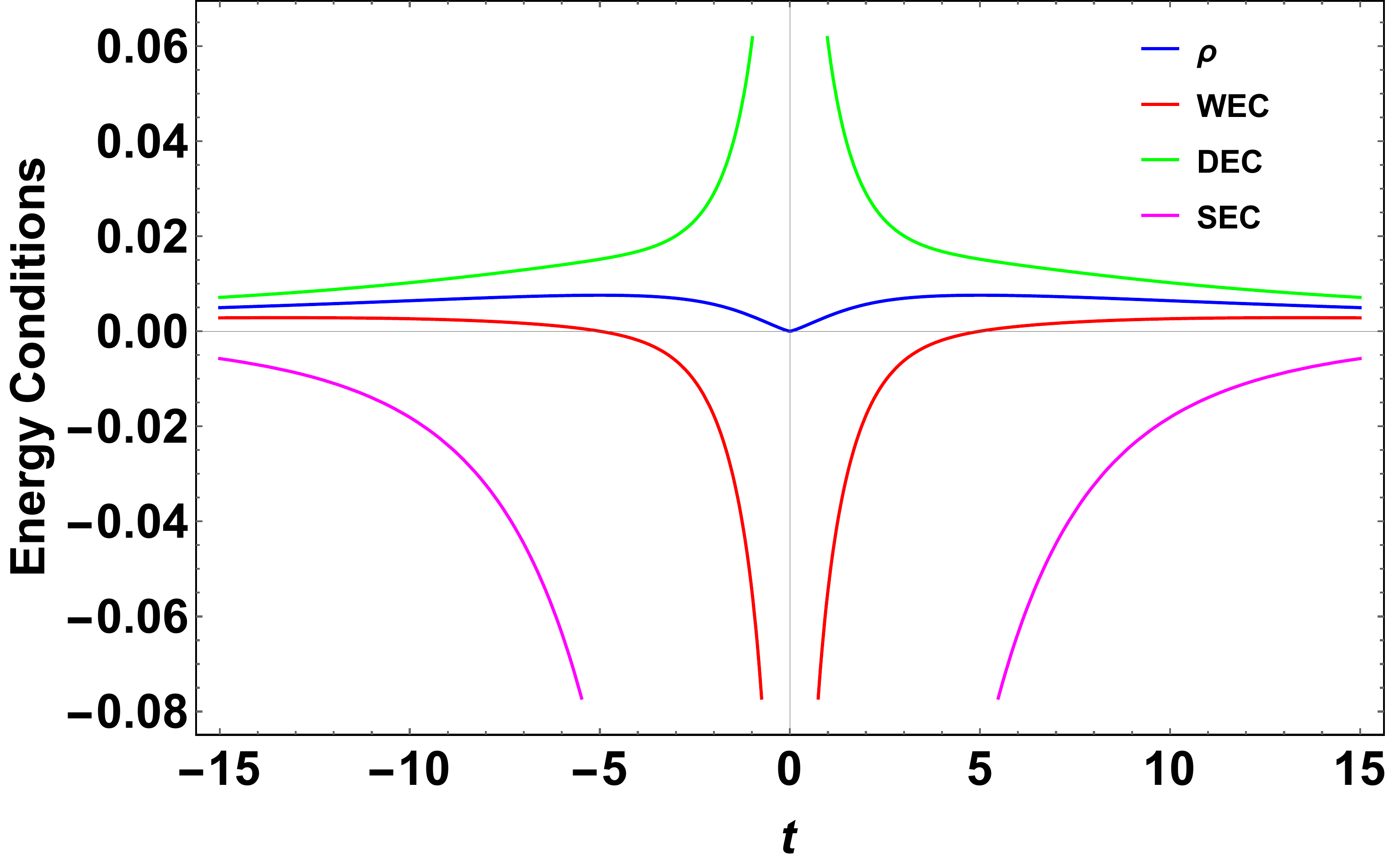}
\caption{Profiles of energy conditions w.r.t. cosmic time $t$ for $m= -1.5, \,\ n=0.662 $ and $b=5$ (for model-2).}
\label{f5}
\end{figure}

\subsection{Model 2: $f(Q)= Q+m Q^n$}

Here, we have considered a non-linear Lagrangian $f(Q)$, i.e. $f(Q)= Q+m Q^n$ to explore the evolution of the universe through a bouncing solution, where $m$ and $n$ are arbitrary constants. The energy-momentum components for the polynomial form of Lagrangian $f(Q)$ are

\begin{equation}
\label{21}
\rho=m \left(-2^{-n-1}\right) 3^n (2 n-1) \left(\frac{t^2}{\left(b^2+t^2\right)^2}\right)^n-\frac{3 t^2}{4 \left(b^2+t^2\right)^2}
\end{equation}
\begin{equation}
\label{18}
\textbf{WEC :}\,\ \rho+p= \frac{1}{3} (b-t) (b+t) \left(\frac{m 2^{1-n} 3^n n (2 n-1) \left(\frac{t^2}{\left(b^2+t^2\right)^2}\right)^n}{t^2}+\frac{3}{\left(b^2+t^2\right)^2}\right)
\end{equation}
\begin{equation}
\label{19}
\textbf{DEC :}\,\ \rho-p=\frac{m 2^{-n} 3^{n-1} (2 n-1) \left(\frac{t^2}{\left(b^2+t^2\right)^2}\right)^n \left((2 n-3) t^2-2 b^2 n\right)}{t^2}-\frac{2 b^2+t^2}{2 \left(b^2+t^2\right)^2}
\end{equation}
\begin{equation}
\label{20}
\textbf{SEC :}\,\ \rho+3\,p-6\,\dot{f_Q}\,H+f= b^2 \left(\frac{m 2^{1-n} 3^n n \left(\frac{t^2}{\left(b^2+t^2\right)^2}\right)^n}{t^2}+\frac{3}{\left(b^2+t^2\right)^2}\right).
\end{equation}

The equation of state parameter $\omega$ for this model can be read as
\begin{equation}
\label{21}
\omega=\frac{p}{\rho}=\frac{2 m 3^n (2 n-1) t^2 \left((4 n-3) t^2-4 b^2 n\right) \left(\frac{t^2}{\left(b^2+t^2\right)^2}\right)^{n-1}+3\ 2^n t^2 \left(t^2-4 b^2\right)}{2 m 3^{n+1} (2 n-1) t^4 \left(\frac{t^2}{\left(b^2+t^2\right)^2}\right)^{n-1}+9\ 2^n t^4}
\end{equation}

\begin{figure}[H]
\centering
\includegraphics[scale=0.4]{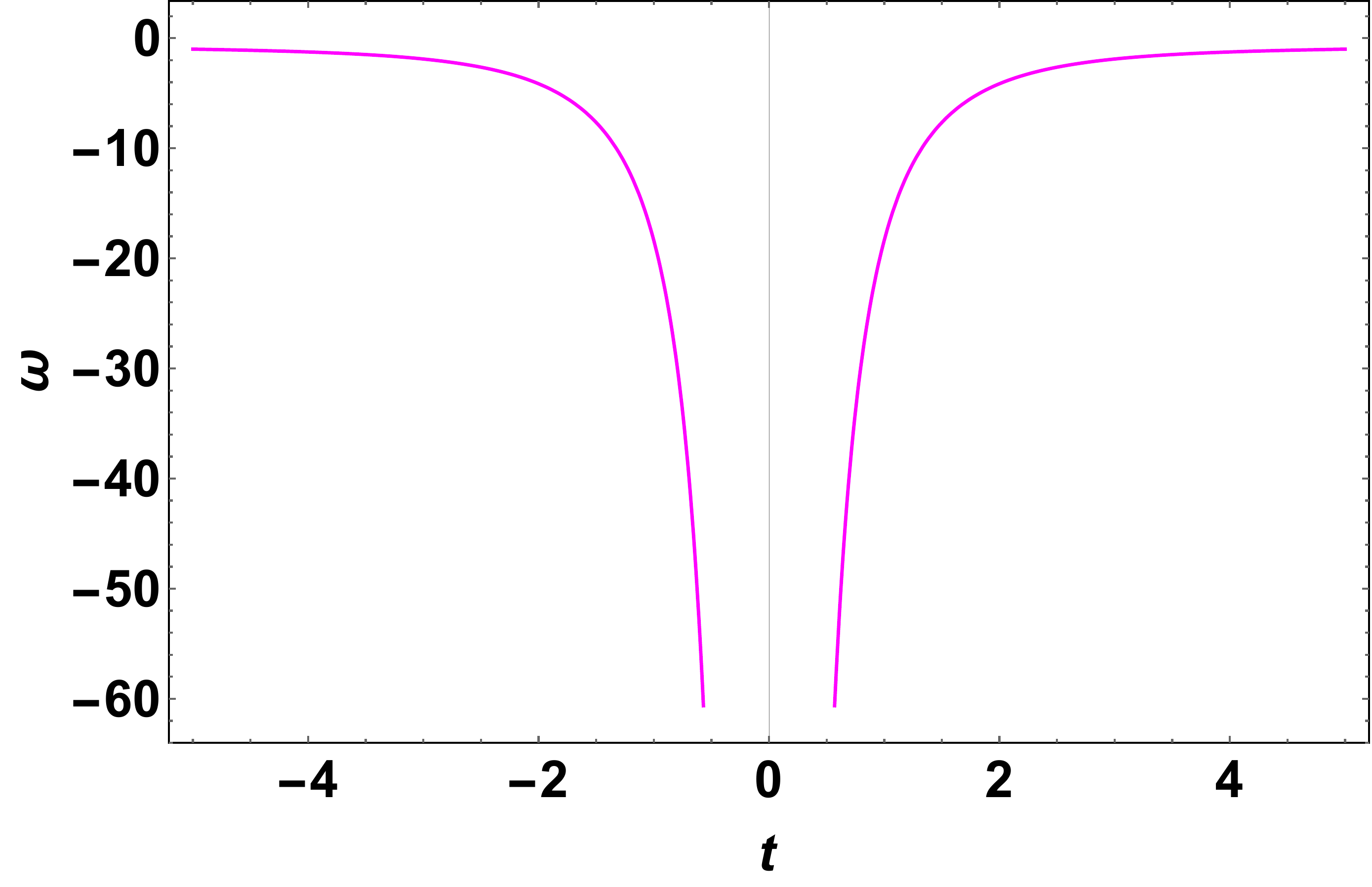}
\caption{Profile of equation of State parameter $\omega$ w.r.t. cosmic time $t$ for $m= -1.5, \,\ n=0.662 $ and $b=5$(for model-2).}
\label{f6}
\end{figure}

Similarly, as we discussed in first model that all of the above expression for energy conditions are depend on the model parameters. Nevertheless, one can not chose the values of parameters randomly, which may violate the energy conditions. Therefore, for second model, we have found that for $n\geq 0.5$ and $m\leq 0$, the energy density will be positive everywhere. The profiles of all energy conditions are depicted in Fig. \ref{f5}, for a particular value of $m,\,\ n$ from their bounded ranges. From the profiles of ECs, we have observed that at the bouncing point $t=0$, NEC and SEC have been violated. Also, the equation of state parameter represents the phantom phase ($\omega<-1$) near the bouncing point, see in Fig. \ref{f6}. Moreover, in literature, it is reported that our universe went through a rapid expansion phase immediately after its' birth. Therefore, it may be the reason for showing our universe's phantom behavior during its early evolution in Fig. \ref{f4} and \ref{f6}.

\section{Perturbation Analysis of Hubble Parameter}\label{VI}
\justifying

In this section, we are interested in investigating the bouncing solution's stability through perturbation analysis in the framework of $f(Q)$ gravity. To do that, first, we have presumed a linear perturbation of the Hubble parameter as
\begin{equation}
\label{22}
H^{*}(t)=H(t)(1+\delta_n(t)),
\end{equation}
where $H^{*}(t)$ is the perturbed Hubble parameter and $\delta(t)$ is the perturbation term.

As we know, the conservation equation energy-matter perfect fluid reads \cite{Lazkoz/2019}

\begin{equation}
\label{23}
\dot{\rho}+3H (\rho+3p)=0.
\end{equation}

Now, in the following, we will evaluate the value of perturbation term and check its' behaviour.

\textbf{\underline{\textit{For model-1}: $f(Q)=\alpha Q$}}

Now, using Eqns. \eqref{22}, \eqref{11}, \eqref{12} and $f(Q)=\alpha Q$ in \eqref{23}, we have calculated the following expression

\begin{equation}\label{24}
\frac{3 \alpha  t \left(4 t \left(b^2+t^2\right) \delta_n'(t)+(\delta_n(t)+1) \left(4 b^2+3 t^2 \delta_n(t)-t^2\right)\right)}{4 \left(b^2+t^2\right)^3}=0
\end{equation}

Solving Eqn. \eqref{24}, we found the perturbation term $\delta_n$ as
\begin{equation}
\label{25}
\delta_n(t)= \frac{4 \left(b^2+t^2\right)}{3 t \left(t-c_1\right)}-1,
\end{equation}
where $c_1$ is the integration constant.

\textbf{\underline{\textit{For model-2}: $f(Q)=Q+m Q^n$}}

\begin{figure}[H]
\centering
\includegraphics[scale=0.4]{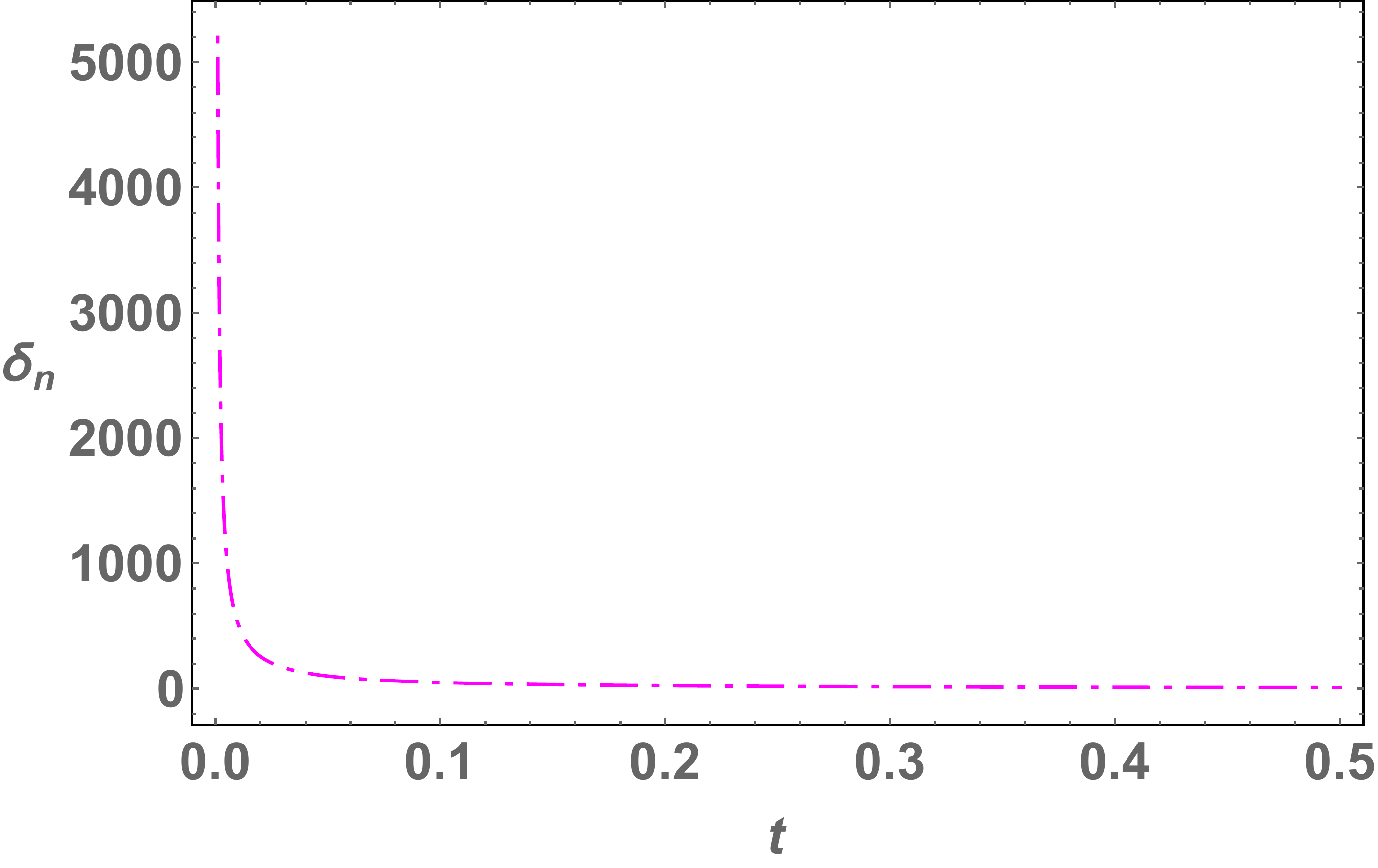}
\caption{Profile of perturbation term $\delta_n$ w.r.t. cosmic time $t$ with $m= -1.5, \,\ n=0.662 $ and $b=5$ for model-2.}
\label{f8}
\end{figure}

For this case, using Eqns. \eqref{22}, \eqref{11}, \eqref{12} and $f(Q)=Q+m Q^n$ in \eqref{23}, we have calculated the following expression
%\begin{widetext}
\begin{multline}\label{26}
\frac{m  2^{1-n } 3^{n } (2 n -1) \left(b^2+t^2\right)^2 \left(\frac{t^2 (\delta_n(t)+1)^2}{\left(b^2+t^2\right)^2}\right)^{n } \left(4 n  t \left(b^2+t^2\right) \delta_n'(t)+(\delta_n(t)+1) \left(4 n  b^2+3 t^2 \delta_n(t)+(3-4 n ) t^2\right)\right)}{4 t \left(b^2+t^2\right)^3 (\delta_n(t)+1)}\\
+\frac{12 t^3 \left(b^2+t^2\right) (\delta_n(t)+1)^2 \delta_n'(t)+12 t^2 \left(b^2+t^2\right) (\delta_n(t)+1)^3+9 t^4 (\delta_n(t)+1)^4-24 t^4 (\delta_n(t)+1)^3}{4 t \left(b^2+t^2\right)^3 (\delta_n(t)+1)}=0
\end{multline}
%\end{widetext}
The above equation is a non-linear differential equation and it is difficult to solve analytically. Therefore, the perturbation term $\delta_n$ evaluated numerically and its' behaviour presented in Fig. \ref{f8}.

\begin{figure}[H]
\centering
\includegraphics[scale=0.4]{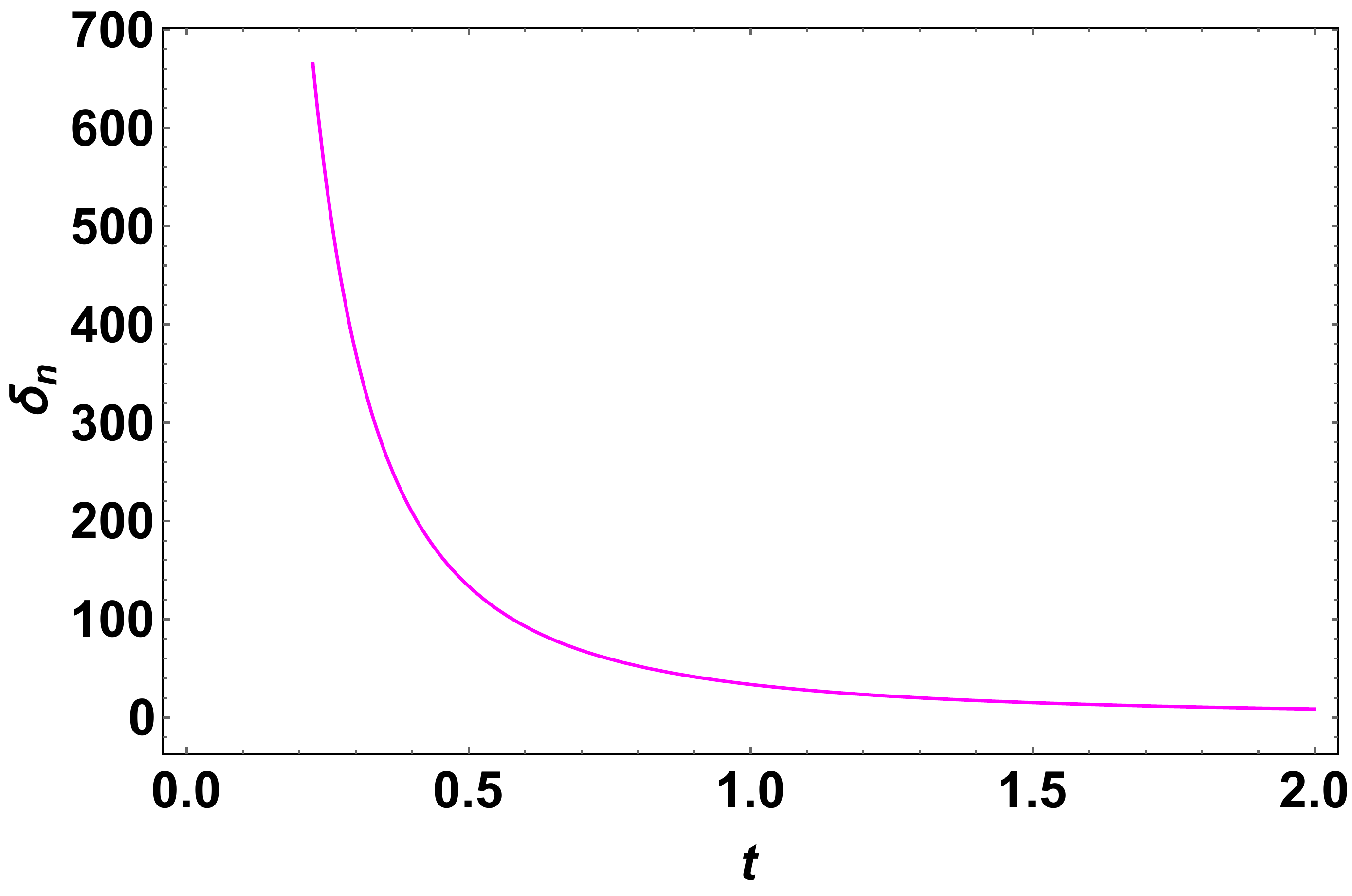}
\caption{Profile of perturbation term $\delta_n$ w.r.t. cosmic time $t$ with $b=5$ for model-1.}
\label{f7}
\end{figure}

In Fig. \ref{f7} and \ref{f8}, profiles of the perturbation term $\delta_n(t)$ shown for model-1 and  model-2, respectively. From those profiles, we have concluded that the value of the perturbed term is very high at the bouncing point and later converges towards zero. From Fig. \ref{f2}, it is observed that at $t=0,\,\ H=0$. Therefore the perturbation term  $\delta_n$ blows up, making the model highly unstable near the bounce. However, away from the bouncing region, the perturbations decay quickly, ensuring stability at late times. We also note that Fig.\ref{f7} and \ref{f8} have valid solutions only for $t >0$. For $t <0$, the functions become imaginary.

\section{Conclusion}\label{VII}

In the last few decades, the singularity problem and inflationary problem questioned the researchers about the origin and evolution of the universe in the absence of sufficient observational data. Therefore, cosmologists follow an attractive approach, i.e., bouncing cosmology to deal with the inflationary paradigm, especially in the absence of initial conditions and absence of initial singularities in the Big Bang model of the universe. In this regard, we have investigated the possibilities of reproducing some of the bouncing models in the newly proposed $f(Q)$ gravity framework, where the non-metricity term $Q$ describes the gravitational interaction. As we know, the linear form of Lagrangian $f(Q)$ mimics the dynamical equations and properties of general relativity (GR). So, for our first bouncing model, we have presumed a linear $f(Q)$ form to retain the fundamental laws of GR. Furthermore, for the second model, we choose a non-linear Lagrangian $f(Q)$.

Moreover, modified gravity is an ideal platform to produce a new cosmological model in which the cosmological barriers are eliminated. In this work, we constructed the cosmological models in the framework of symmetric teleparallel of general relativity (STGR, which is an alternative description of gravity to GR) with FLRW space-time and the presence of perfect fluid matter distribution. In addition, we presumed a bouncing solution to analyse our models.

In the first model, we first derived all the energy conditions and equation of state parameter (EoS) for linear Lagrangian. Besides we checked their profiles in Fig. \ref{f3} and \ref{f4}. From the profile of EoS, we observed that near the bouncing point, it lies in the phantom region ($\omega <-1$). Also, NEC and SEC violated and evolved symmetrically near that point, see Fig. \ref{f3}. It is interesting to note that the SEC has been violated throughout the evolution of cosmic time, suggesting the late-time cosmic acceleration of the universe. Surprisingly, we have found the same kind of results for the non-linear Lagrangian $f(Q)$ (i.e., in model-2) as in model-1.

Finally, we have investigated the stability analysis of the bouncing solution through perturbation analysis. For this, we have considered a linearly perturbed Hubble parameter and used it in our formulation to find out the perturbation effect. The profiles of perturbation tern $\delta_n(t)$ are depicted in Fig. \ref{f7} and \ref{f8} for two models, respectively. From these graphs, one can see that the perturbation effect is more near the bouncing point, and later it reduces to zero. This result suggests that the models are more unstable near the bouncing point, and later it stabilized.

\section*{Acknowledgements} 
S.M. acknowledges Department of Science \& Technology (DST), Govt. of India, New Delhi, for awarding Junior Research Fellowship (File No. DST/INSPIRE Fellowship/2018/IF180676). The work was  supported by the Ministry of Education and Science of the Republic of Kazakhstan, Grant  AP09058240. We are very much grateful to the honorable referee and the editor for the illuminating suggestions that 
have significantly improved our work in terms of research quality and presentation.

\end{document}